\documentclass[sigconf]{acmart}
\usepackage{comment}
\usepackage{fvextra}
\usepackage{tikz}

\usepackage{booktabs}
\usepackage{multirow}
\usepackage{siunitx}

\usepackage{alertmessage}

\usetikzlibrary{arrows.meta,positioning,shapes}

\AtBeginDocument{%
  }

\setcopyright{acmlicensed}
\copyrightyear{2026}
\acmYear{2026}
\acmConference[Conference acronym 'XX]{XX}{2026}{Woodstock, NY}

\begin{document}

\title{TubiFM: Unified Item, Carousel, and Search Ranking for Streaming Discovery}

\author{Alexandre Salle}
\affiliation{%
  \institution{Tubi}
  \city{Porto Alegre}
  \country{Brazil}
}
\email{asalle@tubi.tv}

\author{Chenglei Niu}
\affiliation{%
  \institution{Tubi}
  \city{San Francisco}
  \country{USA}
}
\email{chenglei@tubi.tv}

\author{Suchismit Mahapatra}
\affiliation{%
  \institution{Tubi}
  \city{San Francisco}
  \country{USA}
}
\email{smahapatra@tubi.tv}

\author{Xiaoxiao Chen}
\affiliation{%
  \institution{Tubi}
  \city{San Francisco}
  \country{USA}
}
\email{xchen1@tubi.tv}

\author{Suvash Sedhain}
\affiliation{%
  \institution{Tubi}
  \city{San Francisco}
  \country{USA}
}
\email{ssedhain@tubi.tv}

\author{Yaqi Wang}
\affiliation{%
  \institution{Tubi}
  \city{Beijing}
  \country{China}
}
\email{yaqiwang@tubitv.com}

\author{Shervin Shahryari}
\affiliation{%
  \institution{Tubi}
  \city{San Francisco}
  \country{USA}
}
\email{sshahryari@tubi.tv}

\author{Saurabh Agrawal}
\affiliation{%
  \institution{Tubi}
  \city{San Francisco}
  \country{USA}
}
\email{sagrawal@tubi.tv}

\author{Qiang Chen}
\affiliation{%
  \institution{Tubi}
  \city{San Francisco}
  \country{USA}
}
\email{qiang@tubi.tv}

\author{Michael Tamir}
\affiliation{%
  \institution{Tubi}
  \city{San Francisco}
  \country{USA}
}
\email{mtamir@tubi.tv}

\renewcommand{\shortauthors}{Salle et al.}

\begin{abstract}
Personalized discovery systems often train separate models for item ranking, carousel ranking, and search, even though these tasks expose complementary signals from the same viewer journey: watches shape carousel and item ranking, search queries reveal intent even when they do not lead to a catalog match, and watch history helps interpret search as rewatching, continuation, or new discovery. We introduce the \emph{user story}, a serialized representation that turns a user's cross-surface history---attributes, sessions, watch events with surface and carousel context, and search events---into a single token sequence. By interleaving pretrained language tokens with domain-specific event tokens, user stories let heterogeneous recommendation and search tasks be expressed as prompted next-token prediction over a shared grammar. TubiFM is one instantiation of this approach: a Llama~3.2 1B-based model trained on user stories and prompted to rank items, carousels, or search results without task-specific architectures. In offline evaluation, this single model outperforms specialist baselines across item, carousel, and search ranking. In online A/B tests, TubiFM significantly improves search total viewing time (TVT) by $+3.9\%$ and carousel TVT by $+0.30\%$. Item ranking is statistically neutral on TVT ($+0.14\%$), but matches a mature production stack; across all three tasks, TubiFM serves on L40S GPUs and reduces p99 ranking latency from 500ms to 200ms. These results show that shared user stories can improve discovery while simplifying ranking systems.
\end{abstract}



\maketitle

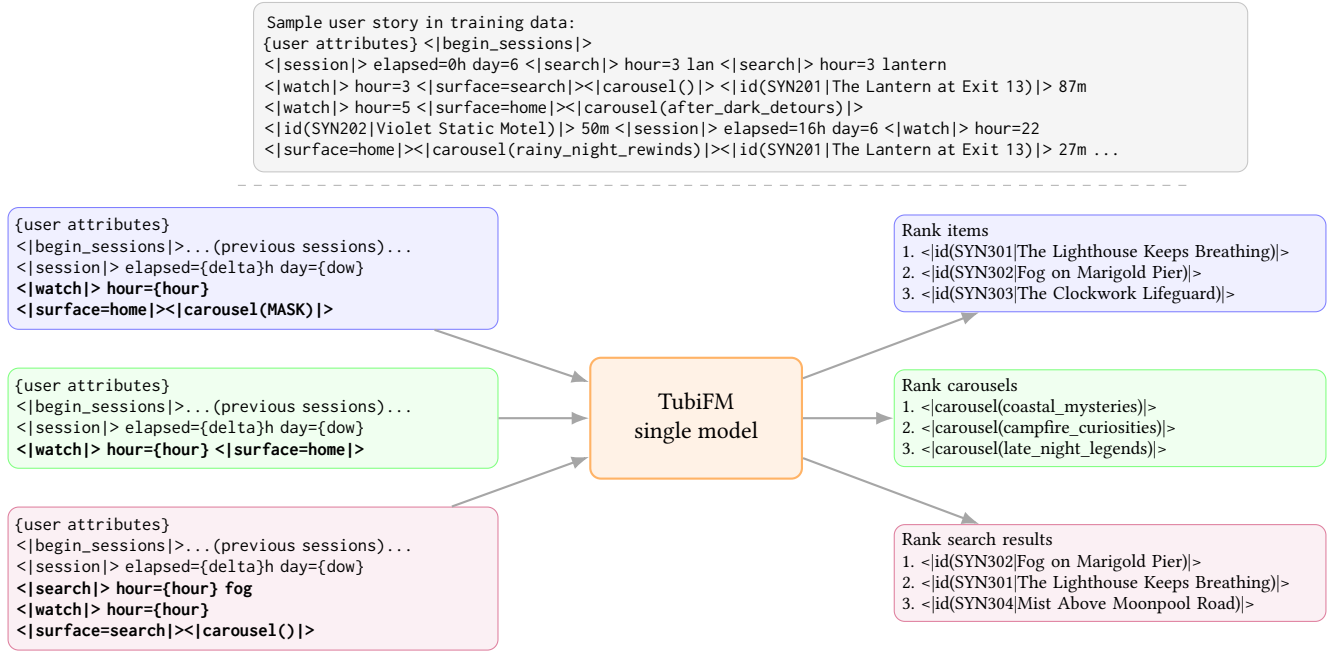
\begin{figure*}[t!]
\centering
\begin{tikzpicture}[
  promptA/.style={draw, rounded corners, align=left, text width=6.2cm, inner sep=4pt, font=\footnotesize\ttfamily, fill=blue!6, draw=blue!50},
  promptB/.style={draw, rounded corners, align=left, text width=6.2cm, inner sep=4pt, font=\footnotesize\ttfamily, fill=green!6, draw=green!50},
  promptC/.style={draw, rounded corners, align=left, text width=6.2cm, inner sep=4pt, font=\footnotesize\ttfamily, fill=purple!6, draw=purple!50},
  trainbox/.style={draw, rounded corners, align=left, text width=12.8cm, inner sep=5pt, font=\footnotesize\ttfamily, fill=gray!8, draw=gray!45},
  sectionlabel/.style={font=\bfseries},
  outA/.style={draw, rounded corners, align=left, text width=5.5cm, inner sep=3pt, font=\footnotesize, fill=blue!6, draw=blue!50},
  outB/.style={draw, rounded corners, align=left, text width=5.5cm, inner sep=3pt, font=\footnotesize, fill=green!6, draw=green!50},
  outC/.style={draw, rounded corners, align=left, text width=5.5cm, inner sep=3pt, font=\footnotesize, fill=purple!6, draw=purple!50},
  model/.style={draw, thick, rounded corners, align=center, minimum width=2.8cm, minimum height=1.6cm, fill=orange!10, draw=orange!60},
  arr/.style={-Latex, thick, draw=gray!70}
]
  \node[sectionlabel, anchor=west] (trainlabel) at (0, 2.4) {Training};
  \node[trainbox, anchor=west] (train) at (0, 1.9) {Sample user story in training data:\\
  \{user attributes\} <|begin\_sessions|>\\
  <|session|> elapsed=0h day=6 <|search|> hour=3 lan <|search|> hour=3 lantern\\ 
  <|watch|> hour=3 <|surface=search|><|carousel()|>
  <|id(SYN201|The Lantern at Exit 13)|> 87m \\<|watch|> hour=5 <|surface=home|><|carousel(after\_dark\_detours)|>\\
  <|id(SYN202|Violet Static Motel)|> 50m <|session|> elapsed=16h day=6 <|watch|> hour=22\\
  <|surface=home|><|carousel(rainy\_night\_rewinds)|><|id(SYN201|The Lantern at Exit 13)|> 27m ...};

  \draw[dashed, gray!60] (-0.2, 0.6) -- (12.4, 0.6);

  \node[promptA] (p1) at (0, -0.5) {\{user attributes\}\\
  <|begin\_sessions|>...(previous sessions)...\\
  <|session|> elapsed=\{delta\}h day=\{dow\}\\
  \textbf{<|watch|> hour=\{hour\} <|surface=home|><|carousel(MASK)|>}};

  \node[promptB, below=0.5cm of p1] (p2) {\{user attributes\}\\
  <|begin\_sessions|>...(previous sessions)...\\
  <|session|> elapsed=\{delta\}h day=\{dow\}\\
  \textbf{<|watch|> hour=\{hour\} <|surface=home|>}};

  \node[promptC, below=0.5cm of p2] (p3) {\{user attributes\}\\
  <|begin\_sessions|>...(previous sessions)...\\
  <|session|> elapsed=\{delta\}h day=\{dow\}\\
  \textbf{<|search|> hour=\{hour\} fog}\\
  \textbf{<|watch|> hour=\{hour\} <|surface=search|><|carousel()|>}};

  \node[model, right=1.2cm of p2] (model) {TubiFM\\single model};

  \node[outA, right=1.2cm of model, yshift=2.05cm] (o1) {Rank items\\
  1.\ <|id(SYN301|The Lighthouse Keeps Breathing)|>\\
  2.\ <|id(SYN302|Fog on Marigold Pier)|>\\
  3.\ <|id(SYN303|The Clockwork Lifeguard)|>};
  \node[outB, right=1.2cm of model] (o2) {Rank carousels\\
  1.\ <|carousel(coastal\_mysteries)|>\\
  2.\ <|carousel(campfire\_curiosities)|>\\
  3.\ <|carousel(late\_night\_legends)|>};
  \node[outC, right=1.2cm of model, yshift=-2.05cm] (o3) {Rank search results\\
  1.\ <|id(SYN302|Fog on Marigold Pier)|>\\
  2.\ <|id(SYN301|The Lighthouse Keeps Breathing)|>\\
  3.\ <|id(SYN304|Mist Above Moonpool Road)|>};

  \draw[arr] (p1) -- (model);
  \draw[arr] (p2) -- (model);
  \draw[arr] (p3) -- (model);

  \draw[arr] (model) -- (o1);
  \draw[arr] (model) -- (o2);
  \draw[arr] (model) -- (o3);
\end{tikzpicture}
\caption{Prompted formulation: changing the prompt switches the ranking task
while keeping a single shared model. In the three inference boxes on the left, bold text marks the
task head appended to the shared story prefix; placeholders such as
\{delta\}, \{dow\} (day of the week), and \{hour\} are filled in at inference
time, corresponding to when predictions are served so results can be
personalized to that moment. The \{user attributes\} placeholder is filled with
ordinary text, not special tokens, and may include categorical or numeric
context such as country and device. This text interface makes arbitrary user
attributes straightforward to add. All viewer journeys and outputs shown are
synthetic illustrative examples. The same serialized story can therefore serve
item, carousel, and search ranking by changing only the prompt suffix.}
\Description{Flow diagram with three prompt variants feeding a single TubiFM
model, which outputs ranked items, carousels, and search results.}
\label{fig:prompting}
\end{figure*}

\section{Introduction}
Personalized discovery in streaming requires modeling the full viewer
journey across watch and search interactions, including the surfaces and
carousels where watches occur. The core ranking tasks in this setting are
therefore coupled rather than independent. A search that fails to produce a
watch can still reveal a genre, franchise, actor, mood, or title the viewer
wanted, which is useful evidence for later item and carousel ranking. Search
also often expresses rewatching or continuation intent, so interpreting a query
depends on the viewer's prior watches. Conversely, the items a viewer watches
provide direct feedback about which carousels and search results should be
ranked higher in future sessions.

In production systems, item ranking, carousel ranking, and search are often
served by separate models, increasing maintenance cost and making these
cross-task signals difficult to exploit. We introduce the user story, a
serialized representation of viewer journeys across surfaces, and present
\emph{TubiFM}, a single generative ranking model that performs all three tasks
within one architecture.
Because this is an industry setting, we use open offline baselines for
reproducible reference points and production A/B tests to measure impact
against Tubi's internal serving systems.
The item-ranking online result is intentionally interpreted as a systems result
rather than a TVT lift: TubiFM is statistically indistinguishable from the
mature production stack on TVT, but it does so with a much simpler treatment
path and substantially lower latency.

\section{Contributions}
\begin{itemize}
\item We propose a hierarchical prompting framework that leverages pretrained LLMs by interleaving natural language
  tokens with domain-specific event tokens, enabling a single production model to
  serve item ranking, carousel ranking, and search ranking tasks from a shared
  user-story interface.
\item We demonstrate the incorporation of sparse textual and
  temporal features into a generative recommendation system, with an
  extensible user-story schema that can readily accommodate additional input
  signals and that suggests a reusable pattern for other recommendation and
  search domains with ordered user events.
\item We evaluate TubiFM against task-specific open baselines and
  task-specific TubiFM variants, and validate the resulting model in
  production A/B tests.
\end{itemize}

\section{Related Work}
Recent work at the intersection of language modeling and recommendation has
developed along several related strands: LLM-based candidate ranking, unified
text-to-text recommendation, generative retrieval over identifiers, and
shared-sequence models of heterogeneous recommendation inputs.

\paragraph{Generative retrieval and semantic identifiers.}
Generative retrieval formulates retrieval and recommendation as sequence
generation rather than nearest-neighbor search, from autoregressive entity
retrieval and differentiable search indices to generative recommenders that
decode structured item identifiers~\cite{decao2020autoregressive,tay2022transformer,mehta2022dsi,rajput2023recommender}. This perspective is important
because it collapses representation learning and retrieval into one model and
provides a natural bridge from language modeling to recommendation. TubiFM
shares the sequence-modeling view, but targets ranking in a production setting:
the model scores candidate item or carousel tokens from a prompt rather than
replacing the entire retrieval stack with decoded identifiers.

\paragraph{LLM-based ranking and reranking.}
Prompted and finetuned LLMs have also been studied as ranking modules for
recommendation~\cite{chatgpt_recsys_2023,palr_2023}. This work shows that
language models can use item text, candidate context, and user history to make
personalized ranking decisions. Recent reranking work further explores whether
explicit reasoning supervision can improve recommendation decisions~\cite{gr2_2026}.
These directions are complementary to our setting. Within this
LLM-as-ranker/reranker literature, the common setup is modular: an upstream
system retrieves candidates and the LLM scores or reranks that set. In contrast,
our user-story representation makes the task itself part of the prompt, allowing
item ranking, carousel ranking, and search ranking to share training data,
vocabulary, and serving code under low-latency production constraints.

\paragraph{Foundation models for recommendation.}
The most directly related direction treats recommenders as unified or
foundation-model-style systems rather than isolated task
models~\cite{geng2022recommendation,zhai2024actions,plum_2025,openonerec_2025}. This
line of work expands beyond prompt design to scale, transfer, tokenization and
identifier design, continued pretraining and alignment, benchmarking, and
deployment. P5 introduced a unified text-to-text framing for recommendation,
HSTU studies large-scale sequential transduction as a closely related
generative-recommender architecture, PLUM emphasizes semantic-ID-based
generative retrieval, and OpenOneRec studies open recommendation foundation
models, holistic benchmarking, and cross-domain transfer.
Diffusion-based work also broadens generative recommendation beyond
autoregressive decoding~\cite{yang2023generate}. TubiFM is complementary to
these systems: our focus is a production discovery setting in which item
ranking, carousel ranking, and search ranking must share signals from the same
viewer journey under low-latency serving constraints.

\paragraph{Unified search and recommendation.}
A growing line of work studies whether search and recommendation can be served
by a single generative model rather than by separate query-driven and
behavior-driven systems. Bridging Search and Recommendation in Generative
Retrieval asks whether one multi-task generative retrieval model can outperform
task-specific search and recommendation models, and shows that joint training
can transfer complementary semantic and collaborative signals across
tasks~\cite{penha2024bridging}. GenSAR makes this trade-off explicit: because
search depends heavily on semantic relevance while recommendation depends more
on collaborative signal, it learns dual-purpose item identifiers with shared
and task-specific codebooks and trains instruction examples for next
recommendation, next search query, next search item, and identifier-language
alignment~\cite{shi2025gensar}. NEO adapts a decoder-only LLM into a
catalog-grounded generator over typed semantic identifiers, enabling
recommendation, text-based retrieval, explanation, and user understanding over
a large heterogeneous catalog~\cite{denadai2026neo}.

TubiFM is aligned with this movement toward unified discovery models, but
differs in the object being unified. Prior systems primarily unify search and
recommendation through identifier design, multi-task prompting or instruction
schemes, and generative retrieval objectives; some also add constrained decoding
or language-steerable output control. TubiFM instead makes the serialized viewer
journey the common interface: a single user story interleaves watch events,
search events, surfaces, carousels, sessions, time, and outcomes. Different
production ranking tasks are exposed by appending task-specific prompt heads to
the same story, supporting item ranking, carousel ranking, and search ranking
with one atomic-token ranker and one low-latency serving path.

\section{User Stories}
\label{sec:user-stories}

Our central data representation is the \emph{user story}: a serialized
account of behavior for one entity over an observation window, written as
an ordered sequence of sessions and events. In streaming discovery, a user
story spans watch and search events, with watch events carrying surface and
carousel context, but the construction is
intended to apply to other recommendation and search domains where user
behavior is naturally observed as a sequence of typed events.

\subsection{Sequence Construction}
Each record is composed of three layers of information: \emph{user attributes},
\emph{session structure}, and \emph{interaction events}.

\paragraph{User attributes.}
Each record begins with an attribute header that may include coarse country,
device context, and other categorical or numeric fields available in a given
domain. These attributes are serialized as ordinary text, which makes it
straightforward to inject additional user attributes without changing the token
grammar.

\paragraph{Session data.}
Viewer activity is organized into sessions. A new session begins when
the viewer has been inactive for more than one hour, and each session
is capped at a maximum duration of twelve hours. Session boundaries are
marked by an explicit \texttt{<|session|>} token, and the elapsed time
since the previous session is encoded as a discrete field, preserving
the temporal rhythm of returning-viewer behavior.

\paragraph{Event data.}
Within each session, two types of interaction events are recorded in
chronological order:
\begin{itemize}
\item \textbf{Watch events}: item identifier, viewing duration,
  originating surface (e.g., home, autoplay, search), carousel (e.g.,
  genre row, editorial collection), and timestamp (day/hour).
\item \textbf{Search events}: query text and timestamp for every search,
  including searches that do not lead to a watch. For search-as-you-type,
  each intermediate query is recorded as a distinct search event (e.g.,
  \texttt{f}, \texttt{fo}, and \texttt{fog}). When a viewer initiates a
  watch from the search results, the watched item is recorded as well.
\end{itemize}

\subsection{Serialization Format}
The three layers above are serialized into a single flat token sequence
per viewer, suitable for autoregressive modeling. The token order is
fixed: attribute header first, then session and event tokens in
chronological order. Figure~\ref{fig:prompting} (top) shows a
representative tokenized journey.

This linearization has two key properties. First, it captures the full
cross-surface trajectory in one sequence, eliminating the need for
task-specific feature engineering or separate interaction logs per
surface. Second, the fixed token grammar enables prompted inference: by
extending the current story with a task-specific prompt head, the same
serialization supports item ranking, carousel ranking, and search ranking
(see Section~\ref{sec:tubifm-model}).

Table~\ref{tab:evaluation-data-stats} summarizes the dataset used in our
streaming setting: an approximately 20M-viewer sample covering watch and search
interactions across four primary surfaces.
\begin{table}[h]
\centering
\footnotesize
\caption{Dataset statistics for the internal streaming sample.}
\label{tab:evaluation-data-stats}
\begin{tabular}{p{0.48\columnwidth}p{0.38\columnwidth}}
\hline
\textbf{Statistic} & \textbf{Value} \\
\hline
Sampled viewers & $\sim$20M \\
Unique titles & $\sim$100K \\
Unique carousels & $\sim$1K \\
Total watches & $\sim$800M \\
Total searches & $\sim$66M \\
Surfaces & Home, Search, Browse, Autoplay \\
Avg.\ events/viewer & $\sim$44 \\
Avg.\ sessions/viewer & $\sim$24 \\
\hline
\end{tabular}
\end{table}

\section{TubiFM Model}
\label{sec:tubifm-model}
TubiFM is a single sequence model trained over the viewer journey with
prompted, autoregressive prediction. Instead of task-specific heads,
different tasks are expressed by changing the prompt and target token
type, allowing one model to serve item ranking, carousel ranking,
and search within the same vocabulary (Figure~\ref{fig:prompting}).

\subsection{Model}
TubiFM is a finetuned Llama~3.2 1B model~\cite{grattafiori2024llama3herdmodels}. We initialize from the public
Llama~3.2 1B checkpoint and continue training on serialized viewer
journeys using autoregressive next-token prediction.
Practically, the model remains a stock text LLM with an extended tokenizer
rather than a custom recommendation architecture. This makes the approach
compatible with the broader LLM tooling ecosystem, including distillation,
quantization, optimized inference packages, and base-model families beyond
Llama.

\subsection{Training Corpora}
\label{sec:training-corpora}
The user-story corpus contains approximately 20M stories with an average
length of roughly 560 tokens, or roughly 11B serialized tokens before repeated
sampling. In addition to behavioral user stories, we build an auxiliary \textbf{catalog
corpus} that maps domain tokens back to text. It contains token-to-text
statements such as \texttt{<|id(\#|title)|> has title \{title\}} and
\texttt{<|carousel(name)|> has name \{name\}}. Inspired by
PLUM~\cite{plum_2025}, this auxiliary objective connects new domain tokens to
the semantic space inherited from Llama pretraining, especially for search,
where lexical query tokens interact with item and carousel identifiers. The training mixture
samples the user-story corpus and the auxiliary catalog corpus in a 20:1 ratio.
All user stories are truncated to 1024 tokens, the context length used
by TubiFM. Offline baselines and task-specific variants are constructed from
these same truncated stories: each method starts from the same serialized
history, and task-specific inputs are formed only by stripping fields from that
history. Thus no baseline receives events outside TubiFM's context window; the
comparison is between models that use the full truncated story and models that
use standard task-specific views derived from it.

\subsection{Training Details}
We train for 120k macro-steps with sequence length 1024, per-device batch size
4, gradient accumulation 4, and 8 GPUs. Training takes approximately 22 hours
on an 8$\times$H100 machine. We use bf16 training, gradient clipping at
1.0, weight decay 0.033, and Adam-style optimization with learning rate
$10^{-5}$ after 1000 warmup steps. Unless stated otherwise, all TubiFM
variants use the same model initialization, sequence length, tokenizer, and
training recipe.

\subsection{Tokenization and Vocabulary}
We interleave the pretrained tokenizer's BPE vocabulary with newly introduced
domain tokens for event types, fields, surfaces, carousels, and item
identifiers. This mixed vocabulary lets the model reuse general language
subwords while treating domain markers and item IDs as atomic units.

\paragraph{Item identifier representation.}
Generative recommenders commonly represent items either as atomic item-ID tokens
or as semantic IDs generated over multiple tokens~\cite{rajput2023recommender,hua2023index}.
Although semantic IDs can reduce item-embedding tables and help with cold start
or catalog churn, those pressures are less central here: Tubi's catalog
contains roughly 100k titles, existing production systems already handle cold
start, and catalog-token refreshes occur roughly monthly.

We therefore represent each item ID as a single atomic token. This choice avoids
fragmenting identifiers across arbitrary subword pieces and makes inference
substantially cheaper: next-token prediction directly produces logits over the
item-token vocabulary, so one forward pass scores the full catalog. By contrast,
semantic-ID top-$K$ retrieval typically requires autoregressive decoding, often
with beam search. Preliminary experiments with semantic IDs did not improve
either offline metrics or online tests, so we use atomic item tokens in all
reported TubiFM results.

\paragraph{Training-time masking.}
We apply stochastic masking during training so the same model can support
container independent ranking and catalog changes between token refreshes. With
probability 0.1, each surface and carousel pair is replaced by
\texttt{<|surface=home|>} followed by \texttt{<|carousel(MASK)|>}. This teaches
the model to score the next item independently of the container in which it
happened to be observed. Intuitively, the masked carousel token trains the
model to approximate a container-marginal score, as if averaging the next-item
prediction over possible carousel identities rather than conditioning on the
observed one. Each content identifier is also replaced with an
unknown item token with probability 0.001, allowing the model to accept new
catalog items at inference time when the catalog has changed since the last
refresh. Masking is disabled at inference time except when the serving prompt
intentionally uses the carousel-mask token for container independent item
ranking.

\paragraph{User attributes and privacy.}
User attributes are coarse, bucketed fields used only inside the internal
training and serving pipeline. We do not release raw user-level records or
attribute values. These attributes are therefore best understood as production
context features rather than public user-profile labels.

\subsection{Tasks}
All tasks are cast as next-token prediction under a task-specific
prompt. The model learns to predict item IDs, carousels, or search
results depending on the prompt and the preceding context, enabling a
single model to generalize across tasks without architectural changes.
At serving time, the prompt extends the existing user story: it either
continues the active session or opens a new one using the same inactivity and
duration rules described above, then appends the task-specific event head.

Figure~\ref{fig:prompting} illustrates the same user story prompted for the
three serving tasks. Item ranking appends a watch-event head, using either a
masked carousel token for container-independent ranking or concrete surface and
carousel tokens for contextual ranking. Carousel ranking appends the next
surface context and predicts a carousel token. Search ranking appends the query
tokens and search-surface watch head, then predicts the item selected from the
results.

\paragraph{Inference scoring.}
At inference time, the serving system constructs the task prompt, runs a
single forward pass, and scores each candidate by the logit of its
corresponding item or carousel token at the next-token position. In the
reported deployment, item and search ranking score the full title vocabulary,
and carousel ranking scores the full carousel vocabulary. Candidate lists can
therefore be ranked without adding a task-specific model head. When a newly
introduced item does not yet have a minted token in the deployed vocabulary, it
can be mapped to the unknown item token until the next catalog-token refresh
introduces a dedicated identifier.

\section{Experiments}

We benchmark TubiFM against task-specific open baselines on data derived from
the same underlying user-story logs. The comparison is between modeling recipes,
not a claim that existing sequential architectures could not be extended with
additional side information. Standard sequential recommenders consume
task-specific event streams; adding search queries, surface context, sessions,
attributes, and multiple targets typically requires task-specific feature
engineering or serving changes. User stories move that integration burden into
the representation: adding a signal is often just adding another tokenized field
or event. Task-specific TubiFM variants provide the cleaner ablation of unified
training because they keep the same backbone and recipe while changing the event
view. Because item ranking, carousel ranking, and search ranking differ in
nature, each task uses its own baseline family.

\subsection{Baselines}

For offline comparisons, we derive task-specific views from the truncated
stories described above while preserving the overall user-story serialization.
The item-ranking view removes search events and carousel information, leaving an
item-centered watch story. The carousel-ranking view removes search events and
item information, leaving a carousel-centered watch story. The search-ranking
view keeps search events and the watched-after-search outcomes, while removing
non-search watches. These views ensure that each task starts from the same
truncated history and differs only in which fields are stripped for that task.

We compare TubiFM against established task-specific methods for each of
the three ranking tasks.

\subsubsection{Item and Carousel Ranking Baselines}
Item and carousel ranking are sequential prediction tasks: predict the next
item or carousel from a viewer's interaction history. We therefore compare
against two state-of-the-art sequential recommendation baselines:
\begin{itemize}
\item \textbf{SASRec}~\cite{kang2018self}: a self-attentive sequential recommendation model
  that captures item-level dependencies through causal self-attention over
  the interaction sequence. It is widely used as a strong next-item prediction
  baseline.
\item \textbf{HSTU}~\cite{zhai2024actions}: a sequential Transformer variant that incorporates
  event-time information through a separate time channel and relative
  attention bias, making it a natural time-aware baseline for streaming
  histories without requiring time and session information to be serialized as
  ordinary input tokens.
\end{itemize}
For item ranking, both models predict the next watched item from the viewer's
watch history; for carousel ranking, the same architectures predict the next
carousel token from the viewer's carousel interaction sequence. Under the
standard sequential recommendation protocol, SASRec consumes only the item or
carousel ID sequence for the target task, while HSTU consumes the same IDs plus
per-watch timestamps through its time channel. These baselines are not designed
to consume TubiFM's full user-story grammar, including search events, surfaces,
sessions, and mixed event types. Both baselines are trained with full softmax
objectives over the same item or carousel vocabulary used for the corresponding
task. To avoid comparing against small sequential models, we evaluated multiple
SASRec and HSTU configurations matched by total parameter scale and wall-clock
training time, and report results for the best configuration on each task.

\paragraph{Task-specific TubiFM variants.}
We also train TubiFM variants with the same initialization and training recipe
but task-specific story views. These variants isolate the effect of multitask
user-story training from the effect of the underlying generative architecture.

\subsubsection{Search Ranking Baselines}
Search ranking differs from the sequential tasks above: the model must
match a query to relevant items rather than predict the next action in a
behavioral sequence. We therefore compare against retrieval-oriented
baselines spanning sparse, zero-shot dense, and supervised dense
approaches:
\begin{itemize}
\item \textbf{BM25}~\cite{robertson2009probabilistic}: a sparse lexical retrieval method that scores items
  by term overlap between the raw query string and title text. BM25 serves as a
  non-neural reference point and remains
  a competitive baseline in many retrieval settings, including standard search
  benchmark settings such as MS MARCO and BEIR~\cite{bajaj2016msmarco,thakur2021beir}.
\item \textbf{Qwen3 Embeddings}~\cite{zhang2025qwen3embeddingadvancingtext}: dense embeddings obtained from Qwen3-4B model
  over title text, used as a zero-shot neural retrieval baseline.
  This tests whether general-purpose language representations can
  capture query--item relevance without domain-specific finetuning.
\item \textbf{Finetuned Sentence-Transformer}~\cite{reimers2019sentencebert}: a
  bi-encoder built on all-MiniLM-L6-v2 and finetuned using Tubi query--title
  positive pairs with in-batch random negatives and sampling bias correction.
  This represents a supervised dense retrieval approach trained on in-domain
  search data.
\end{itemize}

The search labels are derived from watched items following a query, so they
reflect positive engagement rather than exhaustive relevance judgments. This
setup inherits the usual limitations of implicit-feedback search evaluation:
queries can be ambiguous, multiple titles may be relevant, and searches that do
not lead to a watch are ignored rather than converted into negative labels. For
each eligible query, we use the single watched-after-search title as the
positive item. BM25 is therefore a meaningful baseline because the production
systems that generate much of the logged search traffic rely heavily on lexical
matching and because many queries are short prefixes or title-seeking raw
strings. Dense embedding baselines are disadvantaged on very short or partial
queries, where there may be little semantic context to embed; TubiFM can use
the viewer's preceding browse and watch history to disambiguate those sparse
query strings, which is closely related to session-search work that models
query intent through sequential user behavior~\cite{chen2022enhancing}.

\begin{table*}[t!]
\centering
\caption{Main offline results. Bold values are best within each task and metric
and statistically significant at $p < 0.05$.}
\label{tab:results}
\begin{tabular}{llcccccc}
\hline
\textbf{Task} & \textbf{Method} & \textbf{HR@8} & \textbf{HR@50} & \textbf{HR@100} & \textbf{NDCG@8} & \textbf{NDCG@50} & \textbf{NDCG@100} \\
\hline
\multirow{4}{*}{Item Ranking}
& SASRec & 0.3885 & 0.5804 & 0.6566 & 0.2886 & 0.3327 & 0.3451 \\
    & HSTU & 0.4121 & 0.5964 & 0.6640 & 0.3105 & 0.3531 & 0.3640 \\
  & TubiFM task-specific  & 0.4975 & 0.6824 & 0.7524 & 0.395 & 0.4375 & 0.4489 \\
  & TubiFM                & \textbf{0.5817} & \textbf{0.7788} & \textbf{0.8375} & \textbf{0.4599} & \textbf{0.5059} & \textbf{0.5155} \\
\hline
\multirow{4}{*}{Carousel Ranking}
  & SASRec                  & 0.5063 & 0.6097 & 0.6209 & 0.5061 & 0.5289 & 0.5307 \\
  & HSTU                    & 0.7724 & 0.9398 & 0.9760 & 0.5362 & 0.5693 & 0.5732 \\
  & TubiFM task-specific & 0.8138 & 0.9772 & 0.9931 & 0.6214 & 0.6618 & 0.6644 \\
  & TubiFM                & \textbf{0.8343} & \textbf{0.9859} & \textbf{0.9961} & \textbf{0.6366} & \textbf{0.6747} & \textbf{0.6764} \\
\hline
\multirow{5}{*}{Search Ranking}
  & BM25                     & 0.4637 & 0.5193 & 0.7312 & 0.3732 & 0.4112 & 0.4280 \\
  & Qwen3-4B Embeddings        & 0.1365 & 0.1686 & 0.3035 & 0.0946 & 0.1201 & 0.1289 \\
  & Finetuned Sentence-Transformer & 0.1763 & 0.2214 & 0.4091 & 0.1184 & 0.1531 & 0.1659 \\
  & TubiFM task-specific  & 0.5061 & 0.6298 & 0.6758 & 0.4071 & 0.436 & 0.4435 \\
  & TubiFM                & \textbf{0.5673} & \textbf{0.7635} & \textbf{0.8243} & \textbf{0.448} & \textbf{0.4937} & \textbf{0.5036} \\
\hline
\end{tabular}
\end{table*}

\subsection{Evaluation Setup}
Offline experiments use the approximately 20M-viewer sample summarized in
Table~\ref{tab:evaluation-data-stats}. We use a user-level train/evaluation
split: 99\% of users are used for training and 1\% are held out for offline
evaluation. On the held-out users, metrics are computed at every eligible
prediction position rather than only at the final event. For item and search
ranking, eligible positions are watched item-ID tokens under the corresponding
task prompt; for carousel ranking, they are carousel tokens. The context for
each prediction is the story prefix before that token.

For the autoregressive models, offline metrics are computed directly from
next-token probabilities. Given the model-specific context and task prompt, we
run a forward pass and rank tokens by their logits at the target position. For
item and search ranking, the relevant target is the watched item-ID token; for
carousel ranking, it is the carousel token. HR@K is one if the target token
appears among the top $K$ predicted tokens, and NDCG@K discounts the target by
its rank when it appears in the top $K$. TubiFM does not require a separate
retrieval stage for item or search ranking because a single forward pass scores
the full catalog from the shared vocabulary, which contains item IDs, carousel
IDs, event markers, and language tokens. Metrics
are reported at $K \in \{8, 50, 100\}$, matching the cutoffs used for the
task-specific baselines. For offline search ranking, all methods rank the same
candidate universe: the full catalog.

\subsection{Results}
Table~\ref{tab:results} reports the main results across all three tasks. We
evaluate all methods using Hit Rate (HR@K) and Normalized Discounted
Cumulative Gain (NDCG@K)~\cite{jarvelin2002cumulated} at cutoffs
$K \in \{8, 50, 100\}$. Three patterns stand out: TubiFM is the best method on
every metric, the unified model outperforms task-specific TubiFM finetunes, and
the size of the improvement varies by task.

\paragraph{Item ranking.}
TubiFM substantially outperforms the sequential recommendation baselines.
Relative to HSTU, the strongest non-TubiFM baseline, TubiFM improves HR@8 by
41.2\% and NDCG@8 by 48.1\%. The gains remain large at deeper cutoffs, with
HR@100 increasing from 0.6640 to 0.8375. This indicates that user stories do
more than improve the first recommendation: they produce a better ordering
over the broader ranked list.

\paragraph{Carousel ranking.}
Carousel ranking is more saturated in absolute terms, with strong methods
already reaching high HR@50 and HR@100. Even so, TubiFM improves over HSTU by
8.0\% HR@8 and 18.7\% NDCG@8. The larger gain on NDCG@8 matters because
carousel ranking is most sensitive to the first few positions on the home
surface, where exposure is concentrated.

\paragraph{Search ranking.}
Search ranking shows the clearest benefit of combining browsing history and
query tokens in one generative model. BM25 is the strongest non-TubiFM search
baseline, substantially outperforming both dense embedding methods, yet TubiFM
still improves over BM25 by 22.3\% HR@8, 47.0\% HR@50, and 20.0\% NDCG@8. The
large HR@50 gain shows that the watched item appears in the top-50 much more
often, while the NDCG@8 gain shows that the improvement also reaches the
top-ranked positions.

\paragraph{Unified versus task-specific training.}
The unified model also outperforms separate TubiFM finetunes for each task:
by 16.9\% HR@8 on item ranking, 2.5\% HR@8 on carousel ranking, and 12.1\%
HR@8 on search ranking. This is the central result of Table~\ref{tab:results}:
a single user-story model is not merely a simpler serving abstraction, but a
stronger model. Item watches, carousel exposures, and searches provide
complementary views of viewer intent, and modeling them in one sequence
improves ranking quality across both browsing and search surfaces.

\section{Offline Analysis}
Beyond the main benchmark in Table~\ref{tab:results}, we analyze which parts
of the user-story construction drive performance. We focus on ablations over
model initialization and serialized feature groups. We do not report the
internal production rankers in the offline table because those systems combine
multiple proprietary retrieval, filtering, feature, and ranking components;
instead, production systems are used as controls in the online A/B tests.

\subsection{Ablations}
We ablate the main components of the user-story representation to
understand which signals drive ranking quality. Unless otherwise noted,
\emph{vanilla} denotes the full TubiFM configuration: Llama initialization
with user attributes, catalog corpus, session boundaries, temporal fields,
and watch-duration information. In ablation tables, N@K abbreviates NDCG@K.

\paragraph{Initialization.}
Table~\ref{tab:init_ablation} compares continued training from the Llama
checkpoint with training the same architecture from random initialization.
Pretraining improves every metric across all three tasks, with the largest
absolute gains on search and item ranking. This suggests that language
pretraining is useful even though most prediction targets are domain tokens:
the model can reuse pretrained sequence modeling capacity while adapting to
the recommendation vocabulary.

\begin{table}[t]
\centering
\footnotesize
\setlength{\tabcolsep}{3.2pt}
\renewcommand{\arraystretch}{1.02}
\caption{Initialization ablation.}
\label{tab:init_ablation}
\begin{tabular}{l
                S[table-format=1.4]
                S[table-format=1.4]
                S[table-format=1.4]
                S[table-format=1.4]
                S[table-format=1.4]
                S[table-format=1.4]}
\toprule
\textbf{Init} & {\textbf{HR@8}} & {\textbf{HR@50}} & {\textbf{HR@100}} & {\textbf{N@8}} & {\textbf{N@50}} & {\textbf{N@100}} \\
\midrule
\multicolumn{7}{l}{\textbf{Item}} \\
Llama & \textbf{0.5817} & \textbf{0.7788} & \textbf{0.8375} & \textbf{0.4599} & \textbf{0.5059} & \textbf{0.5155} \\
Rand  & 0.5670 & 0.7659 & 0.8249 & 0.4441 & 0.4905 & 0.5001 \\
\addlinespace[1.5pt]

\multicolumn{7}{l}{\textbf{Carousel}} \\
Llama & \textbf{0.8343} & \textbf{0.9859} & \textbf{0.9961} & \textbf{0.6366} & \textbf{0.6747} & \textbf{0.6764} \\
Rand  & 0.8315 & 0.9850 & 0.9959 & 0.6346 & 0.6731 & 0.6749 \\
\addlinespace[1.5pt]

\multicolumn{7}{l}{\textbf{Search}} \\
Llama & \textbf{0.5673} & \textbf{0.7635} & \textbf{0.8243} & \textbf{0.4480} & \textbf{0.4937} & \textbf{0.5036} \\
Rand  & 0.5535 & 0.7480 & 0.8070 & 0.4310 & 0.4763 & 0.4859 \\
\bottomrule
\end{tabular}
\end{table}

\paragraph{User attributes.}
Table~\ref{tab:attribute_ablation} removes coarse user attributes from
the serialized header. The full header is usually best, but the effects are
small across all tasks and metrics. Removing all attributes produces modest
drops, while removing profile attributes or location information changes the
metrics by only a few thousandths; in search, removing location information even
slightly improves HR@8 while reducing the deeper-cutoff and NDCG metrics. We
therefore interpret these fields as weak contextual priors rather than a major
source of model quality.

\begin{table}[t]
\centering
\footnotesize
\setlength{\tabcolsep}{3.2pt}
\renewcommand{\arraystretch}{1.02}
\caption{Coarse attribute ablation.}
\label{tab:attribute_ablation}
\begin{tabular}{lcccccc}
\toprule
\textbf{Variant} & \textbf{HR@8} & \textbf{HR@50} & \textbf{HR@100} & \textbf{N@8} & \textbf{N@50} & \textbf{N@100} \\
\midrule
\multicolumn{7}{l}{\textbf{Item}} \\
vanilla                  & \textbf{0.5817} & \textbf{0.7788} & \textbf{0.8375} & \textbf{0.4599} & \textbf{0.5059} & \textbf{0.5155} \\
excl all info & 0.5773          & 0.7749          & 0.8335          & 0.4558          & 0.5018          & 0.5114          \\
excl profile attrs & 0.5795          & 0.7760          & 0.8344          & 0.4579          & 0.5037          & 0.5132          \\
excl location info & 0.5807          & 0.7774          & 0.8358          & 0.4590          & 0.5048          & 0.5143          \\
\addlinespace[1.5pt]

\multicolumn{7}{l}{\textbf{Carousel}} \\
vanilla                  & \textbf{0.8343} & \textbf{0.9859} & \textbf{0.9961} & \textbf{0.6366} & \textbf{0.6747} & \textbf{0.6764} \\
excl all info & 0.8300          & 0.9843          & 0.9954          & 0.6321          & 0.6708          & 0.6726          \\
excl profile attrs & 0.8308          & 0.9844          & 0.9954          & 0.6334          & 0.6719          & 0.6737          \\
excl location info & 0.8337          & 0.9852          & 0.9958          & 0.6353          & 0.6733          & 0.6751          \\
\addlinespace[1.5pt]

\multicolumn{7}{l}{\textbf{Search}} \\
vanilla                  & 0.5673          & \textbf{0.7635} & \textbf{0.8243} & \textbf{0.4480} & \textbf{0.4937} & \textbf{0.5036} \\
excl all info & 0.5643          & 0.7616          & 0.8197          & 0.4442          & 0.4901          & 0.4995          \\
excl profile attrs & 0.5647          & 0.7625          & 0.8196          & 0.4451          & 0.4911          & 0.5004          \\
excl location info & \textbf{0.5675} & 0.7626          & 0.8227          & 0.4476          & 0.4930          & 0.5028          \\
\bottomrule
\end{tabular}

\end{table}

\paragraph{Session segmentation.}
Table~\ref{tab:session_ablation} removes session segmentation from the
serialized story. In this variant, session delimiters are removed together with
the elapsed-time and day fields attached to each session, leaving a flat
sequence of watch and search events. Removing this structure hurts all tasks,
with the largest relative drops on search and item ranking. Session boundaries
therefore provide useful temporal context beyond the raw order of events, and
we keep them in the full user-story representation.

\begin{table}[t]
\centering
\footnotesize
\setlength{\tabcolsep}{3.2pt}
\renewcommand{\arraystretch}{1.02}
\caption{Session segmentation ablation. \emph{No session} removes session
delimiters and session-level time fields.}
\label{tab:session_ablation}
\begin{tabular}{lcccccc}
\toprule
\textbf{Variant} & \textbf{HR@8} & \textbf{HR@50} & \textbf{HR@100} & \textbf{N@8} & \textbf{N@50} & \textbf{N@100} \\
\midrule
\multicolumn{7}{l}{\textbf{Item}} \\
Full            & \textbf{0.5817} & \textbf{0.7788} & \textbf{0.8375} & \textbf{0.4599} & \textbf{0.5059} & \textbf{0.5155} \\
No session      & 0.5569          & 0.7613          & 0.8254          & 0.4347          & 0.4822          & 0.4926          \\
\addlinespace[1.5pt]

\multicolumn{7}{l}{\textbf{Carousel}} \\
Full            & \textbf{0.8343} & \textbf{0.9859} & \textbf{0.9961} & \textbf{0.6366} & \textbf{0.6747} & \textbf{0.6764} \\
No session      & 0.8285          & 0.9835          & 0.9954          & 0.6257          & 0.6645          & 0.6664          \\
\addlinespace[1.5pt]

\multicolumn{7}{l}{\textbf{Search}} \\
Full            & \textbf{0.5673} & \textbf{0.7635} & \textbf{0.8243} & \textbf{0.4480} & \textbf{0.4937} & \textbf{0.5036} \\
No session      & 0.5410          & 0.7402          & 0.8040          & 0.4204          & 0.4666          & 0.4770          \\
\bottomrule
\end{tabular}

\end{table}

\section{Online Evaluation}
To validate TubiFM against production systems, we run online A/B tests on item
ranking, carousel ranking, and search ranking surfaces. For search and carousel
ranking, TubiFM served live traffic as the online treatment and was adopted for
those surfaces after the experiment. Item ranking is reported as an experiment,
not as an adopted production replacement. Across these tests, the TubiFM
treatment is a single end-to-end ranker in place of multi-stage paths that
combine candidate recall, filtering, feature computation, and separate ranking
stages. This architectural simplification is enabled by the model's ability to
score items directly from the serialized viewer journey without requiring a
dedicated retrieval module.

The online controls are the production systems active at the time of each
experiment. They are stronger but less reproducible than the open offline
baselines because they combine several internal retrieval, filtering, feature,
and ranking components. Each experiment used viewer-level randomization and ran
for one week on millions of users after the standard production ramping process.
The primary metric is total viewing time (TVT). For confidentiality, we report
relative TVT lifts rather than raw traffic counts. Statistical significance is
assessed at $p < 0.05$ using CURE (Control Using Regression Estimates), a
variance-reduction method.

The production systems for all three surfaces have approximately 500ms p99
request-to-ranked-list latency. TubiFM serves on L40S GPUs with dynamic batching
and reduces this p99 latency to approximately 200ms for item, carousel, and
search ranking.

\begin{table}[t]
\centering
\caption{Online A/B test results. Bold TVT lifts are statistically significant
at $p < 0.05$.}
\label{tab:online-results}
\begin{tabular}{lcc}
\toprule
\textbf{Surface} & \textbf{TVT lift} & \textbf{Latency} \\
\midrule
Item & $+0.14\%$ & p99 500ms $\rightarrow$ 200ms \\
\textbf{Search} & \textbf{$+3.9\%$} & p99 500ms $\rightarrow$ 200ms \\
\textbf{Carousel} & \textbf{$+0.30\%$} & p99 500ms $\rightarrow$ 200ms \\
\bottomrule
\end{tabular}
\end{table}

\paragraph{Model refresh.}
For online serving and testing, TubiFM is retrained daily on recent interaction
windows so the active model tracks behavior shifts. Catalog-token refreshes are
less frequent, roughly monthly; newly introduced items and carousels can be
mapped to reserved UNK tokens and then receive dedicated identifiers in a later
catalog-token refresh without architectural changes. In practice, warm-starting
from an existing TubiFM checkpoint lets us train for fewer than 120k macro-steps
while maintaining comparable online performance.

\subsection{Item Ranking A/B Test}
The control for item ranking is the full production recommendation stack:
more than a dozen recallers generate candidates, which are then scored by a
transformer-infused DCN ranker~\cite{wang2017deep}. The treatment evaluates TubiFM as a single
end-to-end alternative to this full pipeline.

The online result is neutral on TVT: TubiFM changes TVT by
$+0.14\%$, which is not statistically significant. We therefore do not
interpret this experiment as evidence that TubiFM is a better item ranker than
the mature production stack. Its value is instead operational: the experimental
treatment omits a large collection of recall and ranking components. In this
setting, matching TVT is useful because it shows that the unified model can
preserve business performance while substantially simplifying serving.

\subsection{Search A/B Test}
The control is a DCN-based~\cite{wang2017deep} two-stage recall-ranking system that consumes the viewer's full watch history along with various embedding features, augmented with NLP matching scores between the query and item metadata. TubiFM serves live traffic as the treatment for this path and yields a statistically significant \textbf{$+3.9\%$
improvement in search TVT} overall relative to this production baseline. The gains are particularly pronounced on tail and long queries, where the model achieves \textbf{+20\% TVT uplift}. This is consistent with the offline observation that TubiFM can use behavioral context when lexical evidence is sparse.

\subsection{Carousel A/B Test}
The control for carousel ranking is a DCN-based model~\cite{wang2017deep} that leverages
the viewer's full watch history and various embedding features to rank
all carousels on the home surface. TubiFM serves live traffic as a single
end-to-end treatment against this production ranker.

TubiFM achieves a statistically significant \textbf{$+0.30\%$ overall TVT lift}
relative to the production baseline. Notably, the gains are concentrated in top-position
carousels, where TubiFM produces a clear increase in TVT, indicating
that the model is more effective at surfacing the most relevant carousel
in the highest-visibility slot. The smaller aggregate lift relative to search
is expected because carousel ranking is already highly optimized and because
most TVT is concentrated in a small number of high-traffic placements.

\section{Limitations}
The offline labels are derived from implicit engagement: item and search ranking
target watched items, while carousel ranking targets the carousel associated
with a watch. These labels are not exhaustive relevance judgments and inherit
the usual biases of logged exposure and positive-only feedback. The online
controls are proprietary systems with internal retrieval, filtering, feature,
and ranking components, so we report relative lifts rather than raw traffic
counts or full serving details. Finally, although user stories are intended as a
general schema, this paper validates them experimentally only in streaming
video.

\section{Conclusion}
This work shows that the central object in recommendation need not be a
surface-specific feature vector, retrieval index, or ranking log, but a
promptable account of the user's journey. By representing browsing,
watching, and searching as one typed temporal sequence, user stories give
TubiFM a shared language for tasks that are usually modeled and served
separately.

The result is both practical and empirical. In offline evaluation on an
approximately 20M-viewer sample, the unified TubiFM model outperforms strong
task-specific baselines and task-specific TubiFM variants across item,
carousel, and search ranking. In production A/B tests, the same modeling
approach improves search and carousel ranking against mature serving
systems, and matches item-ranking TVT in an online experiment while reducing
serving latency across all three surfaces. These results suggest
that the value of recommendation foundation models is not only larger model
capacity, but the ability to place heterogeneous user intent signals into a
common sequence model where they can reinforce one another and simplify
production systems.

More broadly, user stories offer a template for unifying browse and search
in domains where behavior is sequential and hierarchical. Streaming is one
instance of this pattern, but the underlying structure---a user moving
through surfaces, containers, queries, and items over time---appears in
commerce, music, news, and many other discovery products. Treating that
structure as the interface to a foundation model suggests a practical path
toward simpler personalization systems that learn across tasks instead of
fragmenting them.

\bibliographystyle{ACM-Reference-Format}
\bibliography{references}
\end{document}